\documentclass[aps,prb,twocolumn,showpacs,preprintnumbers,amsmath,amssymb,superscriptaddress]{revtex4}

\usepackage{graphicx}% Include figure files
\usepackage{dcolumn}% Align table columns on decimal point
\usepackage{bm}% bold math

\newcommand{\ket}[1]{|#1\rangle}

\begin{document}

\title{Cotunneling through quantum dots coupled to magnetic leads:
zero-bias anomaly for non-collinear magnetic configurations}

\author{Ireneusz Weymann}
\email{weymann@amu.edu.pl} \affiliation{Department of Physics,
Adam Mickiewicz University, 61-614 Pozna\'n, Poland}

\author{J\'ozef Barna\'s}
\affiliation{Department of Physics, Adam Mickiewicz University,
61-614 Pozna\'n, Poland} \affiliation{Institute of Molecular
Physics, Polish Academy of Sciences, 60-179 Pozna\'n, Poland}

\date{\today}

\begin{abstract}
Cotunneling transport through quantum dots weakly coupled to
non-collinearly magnetized leads is analyzed theoretically by
means of the real-time diagrammatic technique. The electric
current, dot occupations, and dot spin are calculated in the
Coulomb blockade regime and for arbitrary magnetic configuration
of the system. It is shown that an effective exchange field
exerted on the dot by ferromagnetic leads can significantly modify
the transport characteristics in non-collinear magnetic
configurations, in particular the zero-bias anomaly found recently
for antiparallel configuration. For asymmetric Anderson model, the
exchange field gives rise to precession of the dot spin, which
leads to a nonmonotonic dependence of the differential conductance
and tunnel magnetoresistance on the angle between magnetic moments
of the leads. An enhanced differential conductance and negative
TMR are found for certain non-collinear configurations.
\end{abstract}

%\keywords{Suggested keywords}

\maketitle

\section{Introduction}

Transport through quantum dots coupled to ferromagnetic leads was
a subject of extensive considerations in the last few years.
\cite{wolf01,loss02,maekawa02,zutic04} Most of the works concerned
theoretical description of spin-polarized transport in the strong
coupling regime, where the Kondo physics emerges,
\cite{surgueevPRB02,lopezPRL03,martinekPRL03,
fransson05,utsumiPRB05,swirkowiczPRB06,eto_simon} as well as in
the weak coupling regime. When the energy of electrons of the
source lead is larger than the corresponding charging energy of
the dot, the current in the weak coupling regime flows through the
system due to sequential (one by one) tunneling {\it via} discrete
levels of the quantum dot. \cite{nato,kouwenhoven97} On the other
hand, if the applied voltage is below threshold for sequential
tunneling, first-order transport is suppressed and the system is
in the Coulomb blockade regime. In this transport regime the
current can still be mediated by cotunneling processes, which
involve correlated tunneling through virtual states of the quantum
dot. \cite{odintsov,nazarov,kang}

Sequential transport through a single-level quantum dot coupled to
ferromagnetic leads was studied for both collinear
\cite{bulka00,rudzinski01} and non-collinear
\cite{exch_field,rudzinski05,wetzels05,braun06} configurations of
the electrodes' magnetic moments. Spin-polarized transport in the
cotunneling regime has been addressed mainly for collinear
systems, \cite{weymannPRB05BR,weymannPRB05,weymannPRB06} although
some aspects of cotunneling through quantum dots coupled to
non-collinearly magnetized leads have also been considered.
\cite{pedersen05,braig05,weymannEPJ05} It has been shown
\cite{weymannPRB05BR} that the interplay of single-barrier and
double-barrier cotunneling processes gives rise to zero-bias
anomaly in the differential conductance $G$ when the leads'
magnetic moments are antiparallel. This, in turn, leads to the
corresponding zero-bias anomaly in the tunnel magnetoresistance
(TMR). Moreover, the even-odd electron number parity effect in TMR
has also been found. \cite{weymannPRB05} On the other hand, if the
dot hosts two orbital levels and the ground state is a triplet,
zero-bias anomaly develops due to cotunneling through singlet and
triplet states of the dot. \cite{weymannEPL06} Very recently,
experimental realizations of quantum dots connected to
ferromagnetic leads have also been presented. Such systems may be
realized for example in granular structures, \cite{zhang05} carbon
nanotubes, \cite{tsukagoshi99,zhao02,jensen05,sahoo05} single
molecules, \cite{pasupathy04} magnetic tunnel junctions,
\cite{fertAPL06} or single self-assembled InAs quantum dots.
\cite{hamaya06}

The main objective of this paper is to address the problem of
spin-polarized cotunneling through a quantum dot coupled to
ferromagnetic leads with non-collinearly aligned magnetic moments.
The external magnetic field required to control magnetic
configuration of the system is assumed to be small enough to
neglect the Zeeman splitting of the dot level. When the system is
in a non-collinear magnetic configuration, an effective exchange
field can significantly modify the transport characteristics.
\cite{exch_field} We show that in the case of symmetric Anderson
model, $\varepsilon=-U/2$, where $\varepsilon$ is the dot level
energy and $U$ denotes the Coulomb correlation parameter, the
zero-bias anomaly in differential conductance decreases
monotonically with increasing deviation of magnetic configuration
from the antiparallel alignment, and eventually vanishes in the
parallel configuration. On the other hand, the effect of exchange
field is most visible when the system is described by an
asymmetric Anderson model, $\varepsilon \neq -U/2$, leading to a
nontrivial variation of the differential conductance with the
angle between magnetizations of the leads.

The anomalous behavior of the differential conductance leads to a
related anomaly in TMR, \cite{weymannPRB05} which for arbitrarily
aligned magnetizations of the leads can be defined as $ {\rm TMR}
= [I_{\rm P} - I(\varphi)]/I(\varphi)$, where $I_{\rm P}$ is the
current flowing through the system in the parallel configuration
$(\varphi=0)$ and $I(\varphi)$ is the current flowing in the
non-collinear magnetic configuration (i.e. when there is an angle
$\varphi$ between the leads' magnetic moments). We show that the
angular dependence of TMR strongly depends on the system
parameters -- in the case of asymmetric Anderson model TMR can
become negative when the leads are aligned non-collinearly.

The paper is organized in the following way. In section
\ref{sec:2} we present the model of single-level quantum dot
coupled to non-collinearly magnetized leads. In section
\ref{sec:3} we briefly describe the real-time diagrammatic
technique used to calculate transport properties. We also present
the perturbation expansion in the Coulomb blockade regime for the
case of arbitrary magnetic configuration of the system. Numerical
results followed by discussion are presented in section
\ref{sec:4}, while the conclusions are given in section
\ref{sec:5}.

\section{Model\label{sec:2}}

\begin{figure}[t]
  \includegraphics[width=0.7\columnwidth]{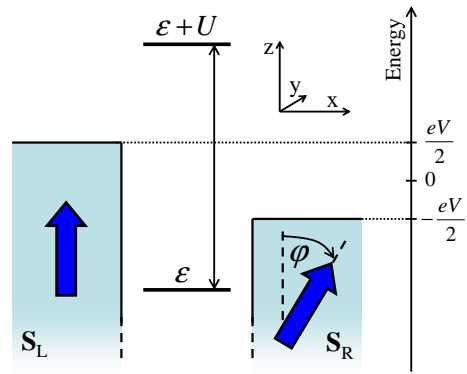}
  \caption{\label{Fig:1}
  (color online) Schematic of a quantum dot coupled to
  ferromagnetic leads with non-collinearly aligned
  magnetizations. The net spin moments of the left
  $\mathbf{S}_{\rm L}$ and right $\mathbf{S}_{\rm R}$ lead
  form an angle $\varphi$. There is a symmetric bias voltage
  applied to the system.}
\end{figure}

The schematic of the system considered is shown in
Fig.~\ref{Fig:1}. The system consists of a single-level quantum
dot coupled to ferromagnetic leads, whose spin moments,
$\mathbf{S}_{\rm L}$ and $\mathbf{S}_{\rm R}$ for left and right
lead, form an arbitrary configuration. The magnetic configuration
of the system is described by the angle $\varphi$ -- when
$\varphi=0$ ($\varphi=\pi$) the system is in the parallel
(antiparallel) magnetic configuration, see Fig.~\ref{Fig:1}.

The system is described by an Anderson-like Hamiltonian of the
form
\begin{equation}
  H = H_{\rm L} + H_{\rm R} + H_{\rm D} + H_{\rm T} \,,
\end{equation}
where the first two components describe noninteracting itinerant
electrons in the leads, $H_{\alpha}=\sum_{k\sigma}
\varepsilon_{\alpha k\sigma}c_{\alpha k\sigma}^\dagger c_{\alpha
k\sigma}$ for the left ($\alpha={\rm L}$) and right ($\alpha={\rm
R}$) lead in the local reference frames. The energy of an electron
with spin $\sigma$ and wave number $k$ in the lead $\alpha$ is
denoted by $\varepsilon_{\alpha k \sigma}$, while $c_{\alpha
k\sigma}^\dagger$ ($c_{\alpha k\sigma}$) is the corresponding
creation (annihilation) operator. The dot is described by $H_{\rm
D}$, which in the global reference frame can be expressed as
$H_{\rm D} =\sum_{\sigma= \uparrow,\downarrow} \varepsilon
d_{\sigma}^{\dagger} d_{\sigma} + U n_{\uparrow}n_{\downarrow}$,
where $\varepsilon$ is the single-particle energy, while
$d_{\sigma}^{\dagger}$ $(d_{\sigma})$ creates (destroys) a
spin-$\sigma$ electron in the dot. The second term of the dot
Hamiltonian describes the Coulomb interaction of two electrons of
opposite spins residing in the dot, with $U$ denoting the
corresponding correlation energy. For the system geometry shown in
Fig.~\ref{Fig:1}, the tunnel Hamiltonian is given by $H_{\rm T} =
H_{\rm TL} + H_{\rm TR}$, with
\begin{equation}
  H_{\rm TL} = \sum_{k} \left(
    T_{{\rm L} +}c_{{\rm L} k+}^{\dagger} d_{\uparrow}
    + T_{{\rm L} -}c_{{\rm L} k-}^{\dagger}d_{\downarrow} +
  {\rm h.c.} \right) \,,
\end{equation}
while $H_{\rm TR}$ reads
\begin{eqnarray}
\lefteqn{  H_{\rm TR} = \sum_{k} \left[\left(
    T_{{\rm R} +}c_{{\rm R} k+}^{\dagger} \cos\frac{\varphi}{2}
    - T_{{\rm R} -}c_{{\rm R} k-}^{\dagger} \sin\frac{\varphi}{2} \right)
    d_{\uparrow}  \right. } \nonumber\\
    &&\left. + \left(
    T_{{\rm R} +}c_{{\rm R} k+}^{\dagger} \sin\frac{\varphi}{2}
    + T_{{\rm R} -}a_{{\rm R} k-}^{\dagger} \cos\frac{\varphi}{2} \right)
    d_{\downarrow} + {\rm h.c.} \right] \,,
\end{eqnarray}
where $T_{\alpha \sigma}$ denotes the tunnel matrix elements
between the dot states and the majority $(\sigma=+)$ or minority
$(\sigma=-)$ electron states in the lead $\alpha$ when
$\varphi=0$. The coupling of the dot to ferromagnetic leads in the
case of collinear magnetic configuration is described by
$\Gamma_{\alpha}^{\sigma}=2\pi |T_{\alpha \sigma}|^2\rho_{\alpha
\sigma}$, where $\rho_{\alpha \sigma}$ is the density of states
for the majority $(\sigma=+)$ and minority $(\sigma=-)$ electrons
in the lead $\alpha$. It is convenient to express the coupling
parameters in terms of spin polarization defined as $p_{\alpha} =
(\Gamma_{\alpha}^{+} - \Gamma_{\alpha}^{-}) / (\Gamma_{\alpha}^{+}
+ \Gamma_{\alpha}^{-})$. Thus, the coupling strength can be
written as $\Gamma_{\alpha}^{\pm}=\Gamma_{\alpha}(1\pm
p_{\alpha})$, where
$\Gamma_{\alpha}=(\Gamma_{\alpha}^{+}+\Gamma_{\alpha}^{-})/2$. In
the following, we assume that the couplings are symmetric,
$\Gamma_{\rm L}=\Gamma_{\rm R} \equiv \Gamma/2$. As reported in
Ref.~\onlinecite{kogan04}, in the weak coupling regime typical
values of the dot-lead coupling strength $\Gamma$ are of the order
of tens of $\mu$eV.

We assume that the magnetic configuration of the system can be
changed by applying a weak external magnetic field. The
corresponding Zeeman energy is then very small and can be
neglected. Thus, following previous theoretical calculations,
\cite{weymannPRB05} we assume that the dot energy level is
spin-degenerate. In the case of quantum dots, the magnetic fields
that produce experimentally measurable effects are rather high,
$\sim 10^4$ Oe. \cite{kogan04}

\section{Method and Transport Equations\label{sec:3}}

In order to calculate transport properties of the system we employ
the real-time diagrammatic technique. \cite{diagrams,koenigdiss}
This technique is based on a systematic perturbation expansion of
the dot density matrix in the dot-lead coupling strength, which
allows one to calculate transport characteristics order by order
in tunneling processes. In the following, we address the problem
of spin-polarized tunneling in the Coulomb blockade regime. In
this transport regime the first-order (sequential) tunneling
processes are exponentially suppressed due to the charging energy
and the current is mediated by second- (cotunneling) or
higher-order tunneling processes. \cite{odintsov,nazarov,kang}

We note that although the approach based on the second-order
perturbation theory and standard rate equation would lead to
reasonable results for the Coulomb blockade regime and collinear
magnetic configurations, it would be insufficient in the case of
arbitrarily aligned magnetizations of the leads. This is because
this approach does not allow for the effects of exchange field. On
the other hand, the advantage of the real-time diagrammatic
technique is that it takes into account the exchange field in a
fully systematic way.

In our considerations we assume that the correlation parameter $U$
between two electrons in the dot is finite, which results in four
dot states $\ket{\chi}$, with $\chi=0,\uparrow,\downarrow,{\rm
d}$, respectively for the empty dot, singly occupied dot with
spin-up or spin-down electron, and for two electrons in the dot.
The density matrix of the dot involves then six elements,
\begin{equation}
  \hat{\rho}_{\rm D} = \left(\begin{array}{cccc}
    P_0^0 & 0 & 0 & 0\\
    0 & P_{\uparrow}^{\uparrow} & P_\downarrow^\uparrow  & 0\\
    0 & P^\downarrow_\uparrow & P_\downarrow^\downarrow  & 0\\
    0 & 0 & 0 & P_{\rm d}^{\rm d}\\
  \end{array} \right) .
\end{equation}
The diagonal elements of the density matrix correspond to the
respective occupation probabilities, while the off-diagonal
elements $P_{\uparrow}^{\downarrow}$ and
$P_{\downarrow}^{\uparrow}$ describe the dot spin $\vec{S}$, with
$S_x={\rm Re}P_\downarrow^\uparrow$, $S_y={\rm
Im}P_\downarrow^\uparrow$, and
$S_z=\left(P_\uparrow^\uparrow-P_\downarrow^\downarrow \right)/2$.
The time evolution of the dot density matrix is described by the
Dyson equation, which can be transformed into a master-like
equation. The master equation allows one to determine the elements
of the density matrix. In the steady state and for spin-degenerate
dot level it can be written as \cite{diagrams,koenigdiss}
\begin{equation} \label{master}
  0 = \sum_{\chi_1^\prime,\chi_2^\prime} P_{\chi^\prime_2}^{\chi^\prime_1}
    \Sigma_{\chi_2^\prime \chi_2}^{\chi_1^\prime \chi_1} \,,
\end{equation}
where $\Sigma_{\chi_2^\prime \chi_2}^{\chi_1^\prime \chi_1}$ is
the irreducible self-energy corresponding to transition forward in
time from state $\ket{\chi_1^\prime}$ to $\ket{\chi_1}$ and then
backward in time from state $\ket{\chi_2}$ to
$\ket{\chi_2^\prime}$.

\subsection{Dot occupations and spin\label{sec:3a}}

We consider transport in the Coulomb blockade regime,
$|\varepsilon|,|\varepsilon+U|\gg \Gamma,k_{\rm B}T$, and when the
dot is singly occupied, $\varepsilon<0<\varepsilon+U$. In order to
calculate the dot occupations and spin, we expand the
self-energies, ${\Sigma_{\chi_2^\prime \chi_2}^{\chi_1^\prime
\chi_1}}=\sum_{j=1}{\Sigma_{\chi_2^\prime \chi_2}^{\chi_1^\prime
\chi_1}}^{(j)}$, and density matrix elements,
${P_{\chi_2}^{\chi_1}}=\sum_{j=0} {P_{\chi_2}^{\chi_1}}^{(j)}$,
order by order in tunneling processes. As a consequence, the
master equation in the first order is given by
\begin{equation} \label{master1}
  0 = \sum_{\chi_1^\prime,\chi_2^\prime} {P_{\chi^\prime_2}^{\chi^\prime_1}}^{(0)}
    {\Sigma_{\chi_2^\prime \chi_2}^{\chi_1^\prime \chi_1}}^{(1)}\, ,
\end{equation}
whereas the second-order master equation reads
\begin{equation} \label{master2}
  0 = \sum_{\chi_1^\prime,\chi_2^\prime} {P_{\chi^\prime_2}^{\chi^\prime_1}}^{(0)}
    {\Sigma_{\chi_2^\prime \chi_2}^{\chi_1^\prime \chi_1}}^{(2)}
    + {P_{\chi^\prime_2}^{\chi^\prime_1}}^{(1)}
    {\Sigma_{\chi_2^\prime \chi_2}^{\chi_1^\prime \chi_1}}^{(1)}
    \,.
\end{equation}
From the above equations and using the normalization,
$\sum_\chi{P_{\chi}^{\chi}}^{(j)}=\delta_{j,0}$, one can find the
density matrix elements in the zeroth and first order. Thus, the
key point is to calculate the first- and second-order
self-energies. This can be done with the aid of the following
rules in the energy space:
\begin{enumerate}
\item Draw all topologically different diagrams with fixed time
ordering and position of vertices. Connect the vertices by
tunneling lines. Assign the energies of respective quantum dot
states to the propagators and a frequency $\omega$ to each
tunneling line.
\item Tunneling lines acquire arrows indicating whether an
electron leaves or enters the dot. For tunneling lines changing
the spin from $\sigma$ to $\sigma^\prime$ assign
$\gamma^{+\sigma\sigma^\prime}_\alpha(\omega)$
$[\gamma^{-\sigma\sigma^\prime}_\alpha(\omega)]$ if the tunneling
line goes backward (forward) with respect to the Keldysh contour
and electron tunnels between the lead $\alpha$ and the dot.
\item For each time interval on the real axis limited by two
adjacent vertices assign a resolvent $1/(\Delta E+i0^+)$, with
$\Delta E$ being the difference of all energies going from right
minus all energies going from left.
\item Each diagram gets a prefactor $(-1)^\gamma$, with $\gamma$
being the number of internal vertices lying on the backward
propagator plus the number of crossings of the tunneling lines,
plus the number of vertices connecting state $\ket{\rm d}$ with
state $\ket{\downarrow}$.
\item Integrate over all frequencies and sum up over the
reservoirs.
\end{enumerate}
The factor $\gamma^{\pm\sigma\sigma^\prime}_\alpha(\omega)$ is
defined as, $\gamma^{\pm\sigma\sigma^\prime}_\alpha(\omega) =
\Gamma^{\sigma\sigma^\prime}_\alpha f^\pm (\omega-\mu_\alpha)$,
with
\begin{eqnarray}
    \Gamma^{\uparrow\downarrow}_{\rm R} &=&
    \frac{1}{4\pi}\sin\varphi \left(\Gamma_{\rm R}^+ - \Gamma_{\rm R}^-\right) \,,\\
    \Gamma^{\uparrow\uparrow}_{\rm R} &=&
    \frac{1}{2\pi}\left(\Gamma_{\rm R}^+ \cos^2\frac{\varphi}{2} +
    \Gamma_{\rm R}^- \sin^2\frac{\varphi}{2}\right) \, ,\\
    \Gamma^{\downarrow\downarrow}_{\rm R} &=&
    \frac{1}{2\pi}\left(\Gamma_{\rm R}^+ \sin^2\frac{\varphi}{2} +
    \Gamma_{\rm R}^- \cos^2\frac{\varphi}{2}\right) \,,
\end{eqnarray}
while $\Gamma^{\sigma\sigma^\prime}_{\rm L}$ is given by changing
${\rm R \leftrightarrow L }$ and setting $\varphi=0$. Here, $f^+$
is the Fermi-Dirac distribution function, and $f^-=1-f^+$. We note
that for arbitrary SU(2) rotations
$\Gamma^{\downarrow\uparrow}_\alpha =
\left(\Gamma^{\uparrow\downarrow}_\alpha\right)^\star$, while for
rotations around the $y$ axis (considered in this paper)
$\Gamma^{\downarrow\uparrow}_\alpha =
\Gamma^{\uparrow\downarrow}_\alpha$.

By using the above rules, we find some useful general relations
for the self-energies which hold in the first and second order
\begin{equation}
    {\Sigma_{\chi \bar{\sigma}}^{\chi\sigma}}^{(j)}
    =-\left[{\Sigma^{\chi\bar{\sigma}}_{\chi\sigma}}^{(j)}\right]^* \,,
\end{equation}
\begin{equation}
    {\Sigma_{\bar{\sigma}\chi}^{\sigma\chi}}^{(j)}
    =-\left[{\Sigma^{\bar{\sigma}\chi}_{\sigma\chi}}^{(j)}\right]^* \,,
\end{equation}
with $\chi=0,\uparrow,\downarrow,{\rm d}$ and $j=1,2$. In
addition,
\begin{equation}
    {\Sigma_{\bar{\sigma}\chi^\prime}^{\sigma\chi}}^{(j)}
    =-\left[{\Sigma^{\bar{\sigma}\chi^\prime}_{\sigma\chi}}^{(j)}\right]^* \,,
\end{equation}
for $\chi,\chi^\prime=\sigma,\bar{\sigma}$
($\chi\neq\chi^\prime$), and
${\Sigma_{\bar{\sigma}\sigma}^{\sigma\bar{\sigma}}}^{(1)}=0$, as
transition between such states has to involve at least two
tunneling lines. In general, the self-energies are complex.
However, as we show in the following, in the case of the Coulomb
blockade regime one only needs to determine the imaginary parts of
the second-order self-energies. This considerably simplifies
numerical calculations, since the self-energies can be determined
analytically.

First of all, we note that in the Coulomb blockade regime the
first-order self-energies associated with energetically forbidden
transitions are exponentially small, i.e., ${\Sigma_{\sigma
\chi}^{\sigma \chi}}^{(1)}=0$, for $\chi=0,\uparrow, \downarrow,
{\rm d}$, and ${\Sigma_{\bar{\sigma} \chi}^{\sigma
\chi}}^{(1)}=0$, for $\chi=0,{\rm d}$. Furthermore, one also has
${\rm Im}{\Sigma_{\sigma \sigma}^{\sigma \bar{\sigma}}}^{(1)}=
{\rm Im}{\Sigma_{\sigma \sigma}^{\bar{\sigma}\sigma}}^{(1)} = {\rm
Im}{\Sigma_{\sigma \sigma}^{\bar{\sigma}\bar{\sigma}}}^{(1)} = 0$.
As a consequence, from the first-order master equation,
Eq.~(\ref{master1}), one finds, ${P_0^0}^{(0)} = {P_{\rm d}^{\rm
d}}^{(0)} = 0$, and $S_y^{(0)} = 0$, whereas the occupations
${P_\uparrow^\uparrow}^{(0)}$, ${P_\downarrow^\downarrow}^{(0)}$,
and $S_x^{(0)}$ are related by the equation
\begin{equation}\label{Eq:CB1}
    {P_\uparrow^\uparrow}^{(0)}
    {\Sigma^{\uparrow\uparrow}_{\uparrow\downarrow}}^{(1)}
    + {P_\downarrow^\downarrow}^{(0)}
    {\Sigma^{\downarrow\uparrow}_{\downarrow\downarrow}}^{(1)}
    + S_x^{(0)} {\Sigma^{\uparrow\uparrow}_{\downarrow\downarrow}}^{(1)} = 0
    \,.
\end{equation}
This equation, however, cannot be solved, as it involves three
unknown variables. To solve for ${P_\uparrow^\uparrow}^{(0)}$,
${P_\downarrow^\downarrow}^{(0)}$, and $S_x^{(0)}$, it is
necessary to include the second-order processes. If the system is
in a collinear configuration (either parallel or antiparallel) the
first two self-energies vanish
${\Sigma^{\uparrow\uparrow}_{\uparrow\downarrow}}^{(1)} =
{\Sigma^{\downarrow\uparrow}_{\downarrow\downarrow}}^{(1)}=0$, and
Eq.~(\ref{Eq:CB1}) yields $S_x^{(0)}=0$. It is also worth noting
that the above equation describes the effect of an exchange field
which is exerted by ferromagnetic leads with non-collinearly
aligned magnetizations. \cite{exch_field} This exchange field
results mainly from the first-order processes. To realize the
meaning of Eq.~(\ref{Eq:CB1}) we rewrite this equation in a more
explicit form (see the formulas given in Appendix A)
\begin{equation}\label{Eq:CB1a}
    2{S_z}^{(0)}\mathcal{B}_{\rm R}^{\uparrow\downarrow} =
    {{S_x}^{(0)}} \left(
      \mathcal{B}_{\rm L}^{\uparrow\uparrow} + \mathcal{B}_{\rm R}^{\uparrow\uparrow}
      - \mathcal{B}_{\rm L}^{\downarrow\downarrow} - \mathcal{B}_{\rm R}^{\downarrow\downarrow}
    \right) \,,
\end{equation}
where
\begin{eqnarray}\label{Eq:exch_field}
    \mathcal{B}^{\sigma\sigma^\prime}_\alpha = \Gamma^{\sigma\sigma^\prime}_\alpha{\rm Re} \left[
      \Psi\left( \frac{1}{2} + i\frac{\varepsilon-\mu_\alpha}{2\pi k_{\rm
      B}T} \right) \right. \nonumber \\
     \left. - \Psi\left( \frac{1}{2} + i\frac{\varepsilon+U-\mu_\alpha}{2\pi k_{\rm
      B}T} \right)
    \right]
\end{eqnarray}
describe the exchange field effects, with $\Psi$ denoting the
digamma function. \cite{exch_field} The exchange field results
from the spin-dependent virtual processes between the dot and
leads. The angular variation of the exchange field is mainly
governed by $\mathcal{B}^{\uparrow\downarrow} _{\rm R} \sim p_{\rm
R} \Gamma_{\rm R} \sin\varphi$ ($\mathcal{B}^{\uparrow\downarrow}
_{\rm L}=0$ in the assumed coordinate system). We note that
$\mathcal{B}^{\uparrow\downarrow} _{\rm R}$ vanishes for collinear
magnetic configurations as well as for nonmagnetic leads. As
follows from Eq.~(\ref{Eq:CB1a}), the exchange field in the first
order gives rise to the rotation of the dot spin in the $x-z$
plane. \cite{mu06} Furthermore, we note that it is a purely
many-body effect, as the exchange field vanishes in the
noninteracting case, $U=0$.

Taking into account the aforementioned results and remarks, one
can set up the second-order master equation, Eq.~(\ref{master2}),
for the Coulomb blockade regime. From the master equation in the
second-order one gets a set of equations which, together with
Eq.~(\ref{Eq:CB1}), allows one to determine
${P_\uparrow^\uparrow}^{(0)}$ and
${P_\downarrow^\downarrow}^{(0)}$, ${P_0^0}^{(1)}$ and ${P_{\rm
d}^{\rm d}}^{(1)}$, and also $S_x^{(0)}$ and $S_y^{(1)}$. The
equations which enable full solution of the problem can be written
in a matrix form as
\begin{widetext}
\begin{equation} \label{masterCB}
   \left(\begin{array}{cccccc}
    0 & {\Sigma^{\uparrow\uparrow}_{\uparrow\downarrow}}^{(1)} &
    {\Sigma^{\downarrow\uparrow}_{\downarrow\downarrow}}^{(1)} & 0
    & {\Sigma^{\uparrow\uparrow}_{\downarrow\downarrow}}^{(1)} & 0 \\
    {\Sigma^{00}_{00}}^{(1)} & {\Sigma^{\uparrow 0}_{\uparrow 0}}^{(2)} &
    {\Sigma^{\downarrow 0}_{\downarrow 0}}^{(2)} & 0 & 2i{\rm Im} {\Sigma^{\uparrow 0}_{\downarrow
    0}}^{(2)}& 0 \\
    {\Sigma^{0\uparrow}_{0\uparrow}}^{(1)} & {\Sigma^{\uparrow \uparrow}_{\uparrow \uparrow}}^{(2)} &
    {\Sigma^{\downarrow \uparrow}_{\downarrow \uparrow}}^{(2)} &
    {\Sigma^{\rm d \uparrow}_{\rm d \uparrow}}^{(1)} &
    2i{\rm Im} {\Sigma^{\uparrow \uparrow}_{\downarrow \uparrow}}^{(2)}&
    2i {\Sigma^{\uparrow \uparrow}_{\downarrow \uparrow}}^{(1)} \\
    {\Sigma^{0\downarrow}_{0\downarrow}}^{(1)} & {\Sigma^{\uparrow \downarrow}_{\uparrow \downarrow}}^{(2)} &
    {\Sigma^{\downarrow \downarrow}_{\downarrow \downarrow}}^{(2)} &
    {\Sigma^{\rm d \downarrow}_{\rm d \downarrow}}^{(1)} &
    2i{\rm Im} {\Sigma^{\uparrow \downarrow}_{\downarrow \downarrow}}^{(2)}&
    2i {\Sigma^{\uparrow \downarrow}_{\downarrow \downarrow}}^{(1)} \\
    {\Sigma^{0\uparrow}_{0\downarrow}}^{(1)} & i{\rm Im}{\Sigma^{\uparrow \uparrow}_{\uparrow \downarrow}}^{(2)} &
    i{\rm Im} {\Sigma^{\downarrow \uparrow}_{\downarrow \downarrow}}^{(2)} &
    {\Sigma^{\rm d \uparrow}_{\rm d \downarrow}}^{(1)} &
    i{\rm Im}\left[
    {\Sigma^{\downarrow \uparrow}_{\uparrow \downarrow}}^{(2)}+{\Sigma^{\uparrow \uparrow}_{\downarrow
    \downarrow}}^{(2)}
    \right] &  i {\Sigma^{\uparrow \uparrow}_{\downarrow
    \downarrow}}^{(1)}\\
    0 & \Gamma & \Gamma & 0 & 0 & 0\\
  \end{array} \right)
  \left(\begin{array}{c}
    {P_0^0}^{(1)}\\
    {P_\uparrow^\uparrow}^{(0)}\\
    {P_\downarrow^\downarrow}^{(0)}\\
    {P_{\rm d}^{\rm d}}^{(1)}\\
    S_x^{(0)}\\
    S_y^{(1)}
  \end{array} \right) =
  \left(\begin{array}{c}
    0\\
    0\\
    0\\
    0\\
    0\\
    \Gamma \\
  \end{array} \right) \, ,
\end{equation}
\end{widetext}
where the last row is due to the normalization condition,
${P_\uparrow^\uparrow}^{(0)} + {P_\downarrow^\downarrow}^{(0)} =
1$. Equation~(\ref{masterCB}) involves only the imaginary parts of
the second-order self-energies which can be calculated
analytically. The formulas for the self-energies appearing in
Eq.~(\ref{masterCB}) and having at the ends the off-diagonal
states are given in Appendix A.

From the second-order master equation one also obtains an
additional equation which involves ${P_\uparrow^\uparrow}^{(1)}$,
${P_\downarrow^\downarrow}^{(1)}$, and $S_x^{(1)}$,
\begin{eqnarray}\label{Eq:CB2}
    0&=&{P_\uparrow^\uparrow}^{(0)}
    {\rm Re}{\Sigma^{\uparrow\uparrow}_{\uparrow\downarrow}}^{(2)}
    + {P_\downarrow^\downarrow}^{(0)}
    {\rm Re}{\Sigma^{\downarrow\uparrow}_{\downarrow\downarrow}}^{(2)}
    + S_x^{(0)}{\rm Re} {\Sigma^{\uparrow\uparrow}_{\downarrow\downarrow}}^{(2)}
    \nonumber\\
    &&+{P_\uparrow^\uparrow}^{(1)}
    {\Sigma^{\uparrow\uparrow}_{\uparrow\downarrow}}^{(1)}
    + {P_\downarrow^\downarrow}^{(1)}
    {\Sigma^{\downarrow\uparrow}_{\downarrow\downarrow}}^{(1)}
    + S_x^{(1)} {\Sigma^{\uparrow\uparrow}_{\downarrow\downarrow}}^{(1)}
    \,.
\end{eqnarray}
This equation is analogous to Eq.~(\ref{Eq:CB1}) in the first
order. To solve it for ${P_\uparrow^\uparrow}^{(1)}$,
${P_\downarrow^\downarrow}^{(1)}$, and $S_x^{(1)}$ one would need
to include the higher-order (third-order) diagrams. In the case of
collinear magnetic configurations,
${\Sigma^{\uparrow\uparrow}_{\uparrow\downarrow}}^{(j)} =
{\Sigma^{\downarrow\uparrow}_{\downarrow\downarrow}}^{(j)} = 0$
for $j=1,2$, and from Eq.~(\ref{Eq:CB2}) one finds $S_x^{(1)}=0$,
while $S_x^{(0)}=0$ from the first order master equation. Thus, if
the magnetizations of the leads are aligned collinearly,
Eq.~(\ref{masterCB}) reduces to the one presented in
Ref.~\onlinecite{weymannPRB05}.

\subsection{Green's functions and the current}

The current flowing through the system can be expressed in terms
of the Green's functions as \cite{meir92}
\begin{eqnarray} \label{current}
   I &=& \frac{ie}{2h}\int d\omega {\rm Tr} \left\{
     \left[\mathbf{\Gamma}_{\rm L} f_{\rm L}^+(\omega)
     - \mathbf{\Gamma}_{\rm R} f_{\rm R}^+(\omega)\right]
     \right.\nonumber\\
     &&\left.\times\left[\mathbf{G^{>}}(\omega) - \mathbf{G^{<}}(\omega) \right] +
     \left( \mathbf{\Gamma}_{\rm L} - \mathbf{\Gamma}_{\rm R}
     \right)\mathbf{G^{<}}(\omega)
    \right\} \,,
\end{eqnarray}
with
\begin{equation}
  \mathbf{G^\lessgtr} = \left(\begin{array}{cc}
    G_{\uparrow\uparrow}^\lessgtr & G_{\uparrow\downarrow}^\lessgtr \\
    G_{\downarrow\uparrow}^\lessgtr & G_{\downarrow\downarrow}^\lessgtr
  \end{array} \right) \, .
\end{equation}
The coupling matrices are given by
\begin{equation}
  \mathbf{\Gamma}_{\rm R} = \Gamma_{\rm R} \left(\begin{array}{cc}
    1 + p_{\rm R} \cos\varphi & p_{\rm R} \sin\varphi \\
    p_{\rm R} \sin\varphi & 1 - p_{\rm R} \cos\varphi
  \end{array} \right) \, ,
\end{equation}
for the right lead and
\begin{equation}
  \mathbf{\Gamma}_{\rm L} = \Gamma_{\rm L} \left(\begin{array}{cc}
    1 + p_{\rm L} & 0 \\
    0 & 1 - p_{\rm L}
  \end{array} \right) \, ,
\end{equation}
for the left lead.
The Green's functions $G_{\sigma\sigma^\prime}^{\lessgtr}(\omega)$
are defined as Fourier transforms of
\begin{eqnarray}
   G_{\sigma\sigma^\prime}^{<}(t) &=& i \langle d_{\sigma^\prime}^\dagger(0) d_\sigma
   (t)\rangle \,,\\
   G_{\sigma\sigma^\prime}^{>}(t) &=& -i \langle d_\sigma (t) d_{\sigma^\prime}^\dagger(0)
   \rangle\,.
\end{eqnarray}

In order to calculate the current in each order we expand the
Green's functions systematically in the coupling strength $\Gamma$
\cite{koenigdiss}
\begin{equation}
   G_{\sigma\sigma^\prime}^{\lessgtr} =
   \sum_{m=0}^{\infty} G_{\sigma\sigma^\prime}^{\lessgtr(m)} =
   \sum_{m=0}^{\infty} \sum_{n=0}^{m}
   G_{\sigma\sigma^\prime}^{\lessgtr(n,m-n)}\,.
\end{equation}
Thus, for the first order one has
$G_{\sigma\sigma^\prime}^{\lessgtr(0)} =
G_{\sigma\sigma^\prime}^{\lessgtr(0,0)}$, whereas in the second
order with respect to $\Gamma$ the Green's functions are given by
$G_{\sigma\sigma^\prime}^{\lessgtr(1)} =
G_{\sigma\sigma^\prime}^{\lessgtr(0,1)} +
G_{\sigma\sigma^\prime}^{\lessgtr(1,0)}$. In the case of Coulomb
blockade regime, the first-order current is exponentially
suppressed and the dominant contribution comes from the
second-order processes. Therefore, one only needs to calculate the
Green's functions in the second order. This can be done using the
following rules in the energy space:
\begin{enumerate}
\item Assign the corresponding external and, in higher orders,
internal vertices to a bare Keldysh contour. Connect the external
vertices by a virtual line and the internal vertices by tunneling
lines. Assign the energies of the respective quantum dot states to
the propagators and a frequency $\omega$ to each tunneling line
and the virtual line. Multiply the diagram with an element of the
reduced density matrix corresponding to the initial state.
\item See rule (2) in calculation of self-energies.
\item See rule (3) in calculation of self-energies.
\item See rule (4) in calculation of self-energies.
\item Integrate over all frequencies except for the virtual line
and sum up over the reservoirs.
\end{enumerate}
For the lesser Green's function an additional minus sign appears,
which is due to reversed direction of the virtual line. The
formulas for the Green's functions in the Coulomb blockade regime
are given in Appendix B.

\section{Results and discussion\label{sec:4}}

In the following we discuss numerical results on electronic
transport through a single-level quantum dot coupled to
non-collinearly magnetized leads. We first present the results for
a symmetric Anderson model, $\varepsilon=-U/2$, then we proceed
with the discussion of an asymmetric Anderson model,
$\varepsilon\neq -U/2$. If $\varepsilon=-U/2$, the exchange field
vanishes and both differential conductance and TMR change
monotonically when going from parallel to antiparallel magnetic
configurations. However, for $\varepsilon \neq -U/2$, the effects
of exchange field become important and lead to nontrivial behavior
of transport characteristics. The symmetric and asymmetric
Anderson models can be realized experimentally by applying a gate
voltage to the dot. With the gate voltage, one can continuously
shift position of the dot level (while $U$ is unaffected) and thus
study the cross-over from symmetric to asymmetric models.

\subsection{Symmetric Anderson model}

\begin{figure}[t]
  \includegraphics[width=0.8\columnwidth]{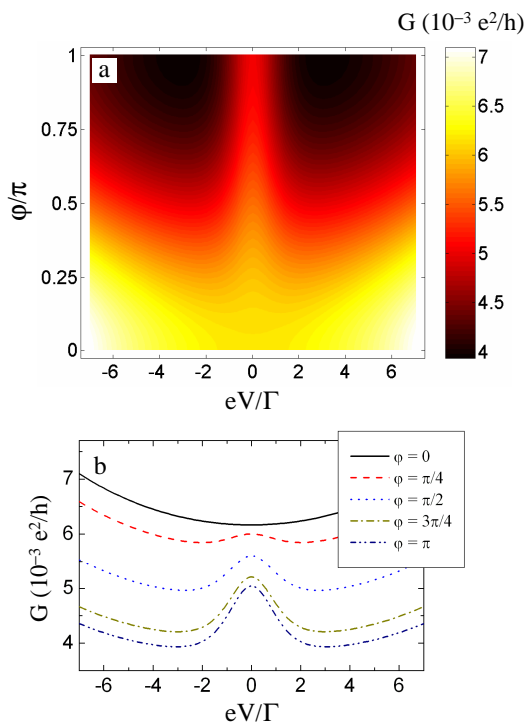}
  \caption{\label{Fig:2} (color online)
  (a) Differential conductance $G=dI/dV$ as a function
  of the bias voltage $V$ and the angle $\varphi$
  for the symmetric Anderson model.
  The parameters are:
  $k_{\rm B} T=0.5 \Gamma$,
  $\varepsilon=-15\Gamma$, $U=30\Gamma$, and
  $p_{\rm L} = p_{\rm R}\equiv p=0.5$.
  Part (b) shows the bias dependence of
  differential conductance for several angles.}
\end{figure}

The differential conductance $G$ as a function of the bias voltage
and the angle $\varphi$ is shown in Fig.~\ref{Fig:2}(a), while
Fig.~\ref{Fig:2}(b) displays the bias dependence of $G$ for
several values of $\varphi$. In the collinear, i.e. parallel and
antiparallel magnetic configurations we recover the earlier
results, \cite{weymannPRB05BR} with the zero-bias anomaly present
in the antiparallel configuration and a parabolic dependence of
$G$ on bias voltage for the parallel configuration. Now, we find a
monotonic variation of the anomaly with the angle between leads'
magnetic moments. When the angle increases from zero to $\pi$, the
anomaly emerges at small values of $\varphi$ and its relative
height increases with increasing angle, reaching a maximum value
at $\varphi =\pi$. The anomaly results from the interplay of
nonequilibrium spin accumulation in the dot and the competition
between the single-barrier and double-barrier cotunneling
processes. The physical mechanism of this anomalous behavior was
discussed in detail in Ref.~[\onlinecite{weymannPRB05BR}],
therefore, we will not describe it here. Instead, we will focus on
its variation with the angle between magnetic moments of the leads
and on the difference between symmetric and asymmetric Anderson
models.

In the linear response regime and for symmetric Anderson model the
exchange field, Eq.~(\ref{Eq:exch_field}), is negligible and the
average dot spin tends to zero. It is therefore relatively easy to
find an approximate formula for the differential conductance. For
$\varepsilon = -U/2$, $p_{\rm L}=p_{\rm R}\equiv p$, and at low
temperatures, the angular dependence of the linear conductance is
given by
\begin{equation}\label{Eq:Glin}
  G = \frac{e^2\Gamma^2}{2h\varepsilon^2}\left[
  3-p^2\left( 1+2\sin^2 \frac{\varphi}{2}\right)
  \right] \,.
\end{equation}
The variation of $G$ with $\varphi$ at low bias is thus
characterized by the factor $1+2\sin^2 (\varphi/2)$, which leads
to maximum (minimum) conductance in the parallel (antiparallel)
magnetic configuration. Such behavior is typical of a normal
spin-valve effect.

\begin{figure}[t]
  \includegraphics[width=0.8\columnwidth]{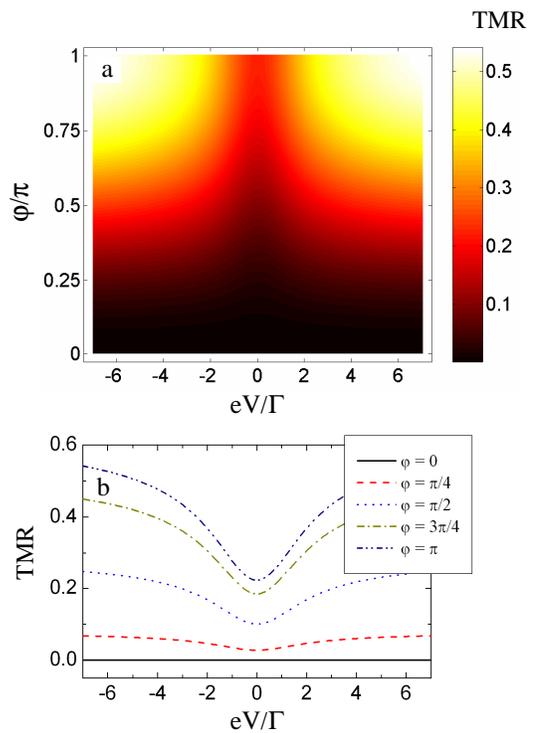}
  \caption{\label{Fig:3} (color online)
   (a) Tunnel magnetoresistance as a function
  of the bias voltage $V$ and the angle $\varphi$
  for the symmetric Anderson model.
  The parameters the same as in Fig.~\ref{Fig:2}.
  Part (b) shows the TMR as a function of the bias voltage
  for several values of $\varphi$.}
\end{figure}

The TMR ratio, defined in the introduction, is shown in
Fig.~\ref{Fig:3}, where part (a) displays the angular and bias
dependence, while part (b) shows the bias dependence of TMR for
several values of the angle $\varphi$. The zero-bias anomaly in
the differential conductance (see Fig.~\ref{Fig:2}) leads to the
corresponding anomaly (dip) in the TMR at small bias voltages. For
collinear configurations we recover again the results presented in
Ref.~[\onlinecite{weymannPRB05}]. The dip in TMR decreases when
magnetic configuration departs from the antiparallel one, and
eventually disappears in the parallel configuration. The variation
of TMR with the angle is monotonic, similarly as the angular
variation of the differential conductance.

With the same assumptions as those made when deriving
Eq.~(\ref{Eq:Glin}), one can approximate the dependence of  TMR on
the angle $\varphi$ at zero bias by
\begin{equation}
  {\rm TMR} = \frac{2p^2 \sin^2 \frac{\varphi}{2}}
  {3-p^2\left( 1+2\sin^2 \frac{\varphi}{2}\right)} \,.
\end{equation}
Now, the angular dependence of TMR is governed by $\sin^2
(\varphi/2)$, which gives maximum TMR in the antiparallel
configuration and zero TMR in the parallel one.

\begin{figure}[t]
  \includegraphics[width=0.7\columnwidth]{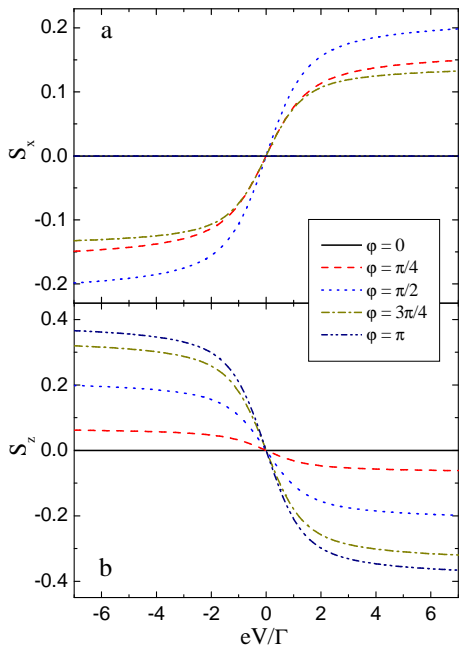}
  \caption{\label{Fig:4} (color online)
  The $S_x$ (a) and $S_z$ (b) components of the dot spin
  as a function of the bias voltage for several values
  of the angle $\varphi$ as indicated in the figure.
  The other parameters are the same as in Fig.~\ref{Fig:2}.
  In part (a) the curves for $\varphi =0$ and $\varphi =\pi$
  coincide.}
\end{figure}

In Fig.~\ref{Fig:4} we show the average components $S_x$ and $S_z$
of the dot spin as a function of the bias voltage, calculated for
several values of the angle $\varphi$. As one can see, $S_x=0$ and
$S_z=0$ in the parallel magnetic configuration ($\varphi=0$),
while in the antiparallel configuration ($\varphi=\pi$) $S_x=0$
whereas $S_z$ is finite. The $y$ component of the dot spin
vanishes, $S_y=0$, irrespective of magnetic configuration of the
system, so it is not shown in Fig.~\ref{Fig:4}. As a consequence,
up to the second order in tunneling processes, the average dot
spin is in the $x-z$ plane, which is a characteristic feature of
the symmetric Anderson model, $\varepsilon=-U/2$. In other words,
there is no precession of the spin out of the $x-z$ plane, that is
out of the plane formed by magnetic moments of the two electrodes.
This is due to the cancellation of different contributions to the
exchange field in the case of symmetric Anderson model. The
absence of exchange field is responsible for the monotonic angular
variation of both the differential conductance and TMR, discussed
above, and also for vanishing of $S_y$ component of the dot spin.

\subsection{Asymmetric Anderson model}

\begin{figure}[b]
  \includegraphics[width=0.8\columnwidth]{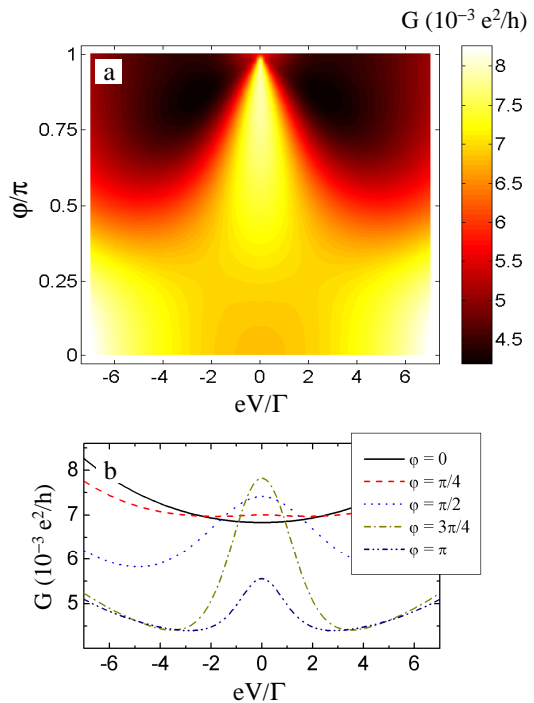}
  \caption{\label{Fig:5} (color online)
  (a) The differential conductance  $G=dI/dV$ as a function
  of the bias voltage $V$ and the angle $\varphi$
  for the asymmetric Anderson model
  $\varepsilon=-12\Gamma$ and $U=30\Gamma$.
  The other parameters are the same as in Fig.~\ref{Fig:2}.
  Part (b) shows the bias dependence of
  differential conductance for several values of the angle  $\varphi$.}
\end{figure}

The transport characteristics described above for the symmetric
Anderson model are significantly modified when the model becomes
asymmetric, i.e. when $\varepsilon \neq -U/2$. This can be
realized by shifting the dot level position by a gate voltage
applied to the dot. Different contributions to the exchange field
do not cancel then, and the resulting effective exchange field
becomes nonzero. This has a significant impact on transport
properties, as we show and discuss below.

First, we note that the leading contribution to the exchange field
comes from the first-order diagrams, whereas the current flows due
to the second-order tunneling processes. This leads to two
different time scales which determine transport characteristics,
as will be discussed  below in more details. Transport properties
are thus a result of the interplay between the first- and
second-order processes. We emphasize that although the first-order
processes do not contribute directly to the current, they
influence transport {\it via} modification of the dot spin.

The strength of effective exchange field is determined by
deviation of the Anderson model from the symmetric one, described
quantitatively by $2\varepsilon+U$ (with $2\varepsilon+U=0$ for
the symmetric model). On the other hand, the cotunneling processes
contributing to the current are proportional to the bias voltage.
Thus, to parameterize relative magnitude of the two processes we
define a dimensionless parameter $x=(2\varepsilon+U)/|eV|$.

\begin{figure}[t]
  \includegraphics[width=0.8\columnwidth]{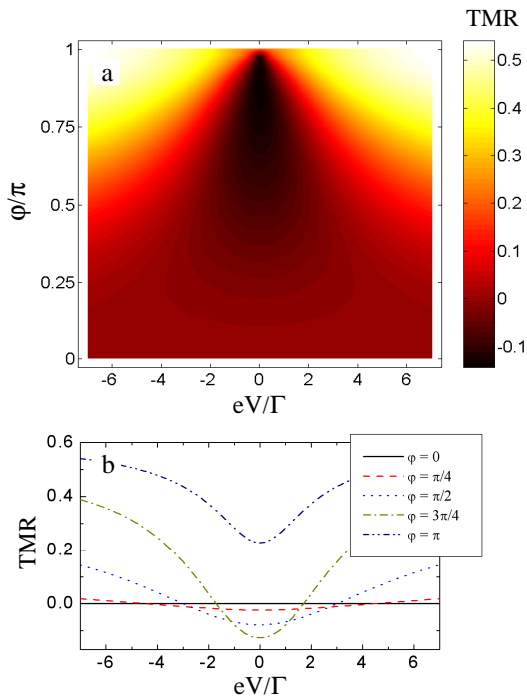}
  \caption{\label{Fig:6} (color online)
  (a) Tunnel magnetoresistance  as a function
  of the bias voltage $V$ and the angle $\varphi$
  for the asymmetric Anderson model.
  The parameters the same as in Fig.~\ref{Fig:5}.
  Part (b) shows the TMR as a function of the bias voltage
  for several values of $\varphi$.}
\end{figure}

In Fig.~\ref{Fig:5}a we show the differential conductance as a
function of angle and bias voltage for an asymmetric Anderson
model, $\varepsilon\neq -U/2$, while Fig.~\ref{Fig:5}b displays
the bias dependence of $G$ for different values of $\varphi$. As
one can see, the low-bias differential conductance becomes
enhanced in a certain range of the angle $\varphi$, leading to a
nonmonotonic dependence of the conductance $G$ on the angle
between the leads' magnetizations. This nonmonotonic behavior is
also visible in TMR. The density plot of TMR is shown in
Fig.~\ref{Fig:6}a, whereas the TMR as a function of bias voltage
for different values of $\varphi$ is shown in Fig.~\ref{Fig:6}b.
There is a range of the angle $\varphi$ between magnetic moments
of the leads, where TMR changes sign and becomes negative in the
small bias region, i.e. the corresponding conductance is larger
than that in the parallel configuration. To discuss and account
for the influence of exchange field, we further present
one-dimensional figures displaying explicitly the angular
dependence of both $G$ and TMR.

\begin{figure}[t]
  \includegraphics[width=0.7\columnwidth]{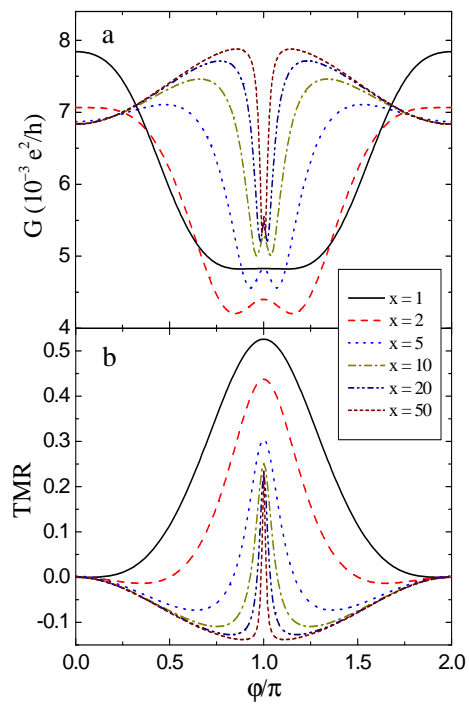}
  \caption{\label{Fig:7} (color online)
  The differential conductance $G$ (a)
  and tunnel magnetoresistance (b)
  as a function of the angle $\varphi$
  for several values of $x=(2\varepsilon+U)/|eV|$
  and for $\varepsilon=-12\Gamma$.
  The other parameters are the same as in Fig.~\ref{Fig:2}.}
\end{figure}

The angular variation of the differential conductance and TMR for
an asymmetric Anderson model is shown in Fig.~\ref{Fig:7} for
several values of the parameter $x$, $x=(2\varepsilon+U)/|eV|$. In
this figure both $\varepsilon$ and $U$ are kept constant
($\varepsilon=-2U/5$ and $U=30\Gamma$), so the variation of $x$ is
only due to the change of the bias voltage. When increasing $x$
(decreasing bias voltage), one increases the relative role of the
first-order processes (leading to the spin precession), which
considerably modifies the angular dependence of the differential
conductance and TMR. As one can see in Fig.~\ref{Fig:7}a, the
differential conductance for $x\gg 1$ does not change
monotonically with the angle $\varphi$. It increases first with
increasing $\varphi$, reaches a local maximum, and then drops to a
local minimum for $\varphi$ close to the antiparallel
configuration, and again slightly increases when approaching
$\varphi=\pi$. Thus, the maximum and minimum differential
conductance occur for non-collinear configurations, and not for
parallel and antiparallel ones. The local maximum in $G$ at the
antiparallel configuration becomes smaller when the effect of
exchange field is suppressed ($x$ decreases) and eventually
disappears for $x=1$ by transforming into the global minimum of
$G$ in the antiparallel configuration. Similarly, the maximum at a
non-collinear configuration disappears with decreasing $x$ and
changes into a global maximum in the parallel configuration.

The most characteristic features of transport characteristics in
the presence of exchange field are the enhanced differential
conductance at a non-collinear alignment and its rapid drop when
approaching the antiparallel configuration. This is most visible
in the curve for $x=50$ in Fig.~\ref{Fig:7}a. In this case the
differential conductance increases until the configuration becomes
close to the antiparallel one, and then drops rapidly to the value
of $G$ in the antiparallel configuration. The key role in this
behavior is played by the first-order processes giving rise to the
exchange field. These processes lead to the precession of spin in
the dot, which facilitates tunneling processes and leads to an
increase in the conductance as compared to the parallel
configuration. When the configuration becomes close to the
antiparallel one, the first-order processes become suppressed and
the conductance drops to that for antiparallel alignment. For
smaller $x$ the drop in conductance begins at a smaller angle,
which follows from the fact that now the relative role of exchange
field decreases due to increased cotunneling rates.

The nonmonotonic behavior of the differential conductance with
$\varphi$ leads to a nonmonotonic dependence of TMR. This is shown
in Fig.~\ref{Fig:7}b for several values of $x$. The effects due to
exchange field give rise to a local minimum in TMR at a
non-collinear magnetic configuration. Moreover, in this transport
regime TMR changes sign and becomes negative. When the magnetic
configuration is close to the antiparallel one, TMR starts to
increase rapidly reaching maximum for $\varphi=\pi$. The negative
TMR and its sudden increase when the configuration tends to the
antiparallel one are a consequence of the processes leading to
nonmonotonic behavior of the differential conductance, as
described above.

\begin{figure}[t]
  \includegraphics[width=0.7\columnwidth]{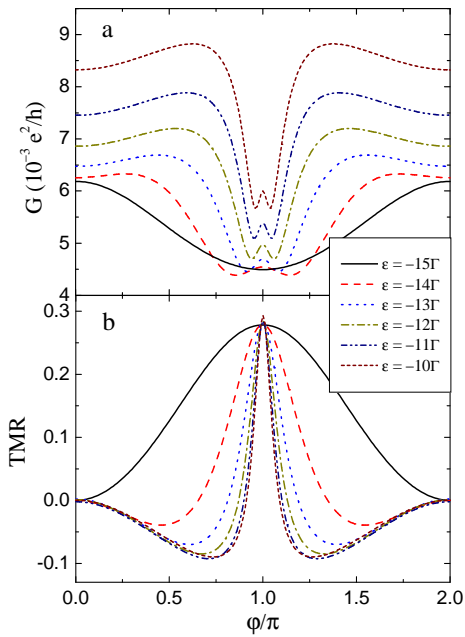}
  \caption{\label{Fig:8} (color online)
  The differential conductance (a) and tunnel magnetoresistance (b)
  as a function of angle between the leads' magnetizations
  for several values of the level position $\varepsilon$,
  as indicated in the figure, and for $V=\Gamma$.
  The other parameters are the same as in Fig.~\ref{Fig:2}.}
\end{figure}

To understand more intuitively the above presented behavior of the
differential conductance and TMR at low bias voltage and close to
the antiparallel configuration, one should consider two different
time scales. One time scale, $\tau_{\rm prec}$, is established by
the virtual first-order processes responsible for the spin
precession due to exchange field,
\begin{equation} \label{Eq:prec_rate}
  \left|\tau^{-1}_{\rm prec}\right| \approx \frac{\Gamma}{2h} p
  \sin\varphi \ln \left|\frac{\varepsilon}{\varepsilon+U}
  \right|\,.
\end{equation}
The second time scale, $\tau_{\rm cot}$, is associated with
second-order processes which drive the current through the system.
At low temperature and bias voltage, the cotunneling rate can be
expressed as
\begin{equation} \label{Eq:cot_rate}
  \tau^{-1}_{\rm cot} \approx \frac{\Gamma^2}{4h}
  (1+p)(1-p\cos\varphi)
  \frac{|eV|U^2}{\varepsilon^2(\varepsilon+U)^2}\,.
\end{equation}
The rate $\tau_{\rm cot}$ depends linearly on the applied voltage,
whereas $\tau_{\rm prec}$ is rather independent of $V$. As a
consequence, at low bias and for non-collinear configuration, the
exchange field plays an important role leading to a nonmonotonic
dependence of differential conductance on the angle between the
leads' magnetizations. When magnetic configuration is close to the
antiparallel one, the spin precession rate is deceased
($\left|\tau^{-1}_{\rm prec}\right|\sim\sin\varphi$) and, at
certain angle, the rate of spin precession becomes comparable to
the cotunneling rate. This gives rise to a sudden drop (increase)
in differential conductance (TMR), as can be seen in
Fig.~\ref{Fig:7}a(b). We note that the nonmonotonic dependence of
differential conductance and magnetoresistance has also been
observed in quantum dots in the strong coupled limit.
\cite{fransson05}

\begin{figure}[t]
  \includegraphics[width=0.7\columnwidth]{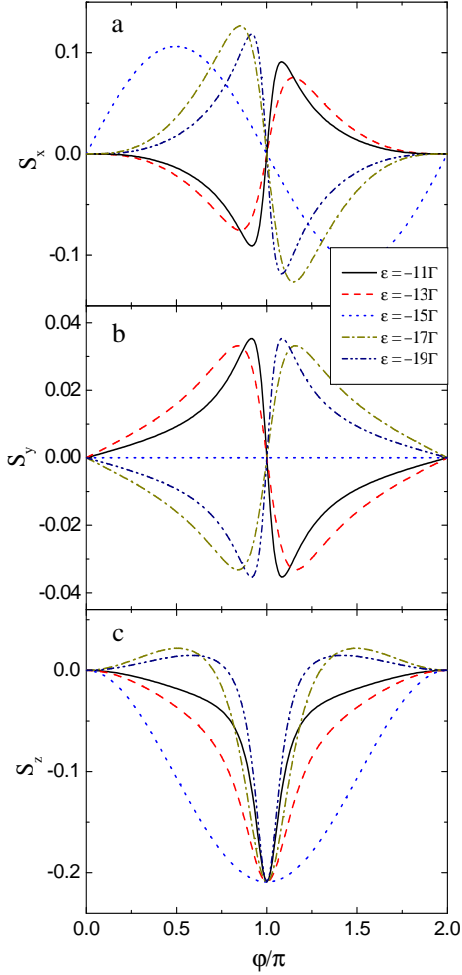}
  \caption{\label{Fig:9} (color online)
  The $S_x$ (a), $S_y$ (b), and $S_z$ (c)
  components of the dot spin as a function of
  angle $\varphi$ for several values of $\varepsilon$ and for $V=\Gamma$.
  The other parameters are the same as in Fig.~\ref{Fig:2}.}
\end{figure}

As follows from Eqs.~(\ref{Eq:prec_rate}) and (\ref{Eq:cot_rate}),
in the low bias voltage regime the influence of the exchange field
is mainly determined by the magnitude of departure from the
symmetric Anderson model, $\left|\tau^{-1}_{\rm prec}\right| \sim
\ln \left|\varepsilon/(\varepsilon+U) \right|$. The angular
variation of the differential conductance and TMR for several
values of the level position is shown in Fig.~\ref{Fig:8}. When
raising the position of the dot level one goes from the symmetric
to asymmetric Anderson model, increasing the effect of exchange
field, which is most visible for $\varepsilon = -10\Gamma$. The
influence of exchange field is the same as in the case of
Fig.~\ref{Fig:7} discussed above. Figure~\ref{Fig:8} demonstrates
that at a certain non-collinear magnetic configuration, by
sweeping the gate voltage one can change the differential
conductance and also tune the TMR between the positive and
negative values. This can be of some importance for future
applications in spintronics.

In the case of symmetric Anderson model the dot spin is located in
the plane defined by the magnetizations of the leads, i.e. in the
$x-z$ plane. On the other hand, in the case of asymmetric Anderson
model, the exchange field gives rise to the spin precession in the
dot, leading to a nonzero $S_y$ component of the dot spin. The
components of the spin in the dot are shown in Fig.~\ref{Fig:9}
for different position of the dot level. If we start from a
symmetric Anderson model, then the conductance and TMR are
independent of the sign of the dot level shift (shifting the dot
level up or down by the same amount leads to equal conductance and
TMR). The spin in the dot, however, depends on the sign of the
shift. This is due to the fact that changing sign of the shift is
associated with a sign change of the exchange field, which is
proportional to $\ln|\varepsilon/(\varepsilon+U)|$. As can be seen
in Fig.~\ref{Fig:9}, the sign of the exchange field determines the
sign of the spin components. It is also worth noting that the two
components, $S_x$ and $S_y$, disappear in both collinear
configurations (parallel and antiparallel). In turn the $S_z$
component vanishes only in the parallel configuration, while it
remains nonzero in the antiparallel configuration due to spin
accumulation. \cite{weymannPRB06}

\section{Summary and concluding remarks\label{sec:5}}

We have considered the cotunneling transport through a
single-level quantum dot coupled to ferromagnetic leads with
non-collinearly aligned magnetic moments. Transport properties
have been calculated with the aid of the real-time diagrammatic
technique which has been adapted to the case of deep Coulomb
blockade regime. The advantage of the real-time diagrammatic
technique is that it takes into account the exchange field
resulting from virtual processes between the dot and ferromagnetic
leads in a fully systematic way.

In the case of symmetric Anderson model, we have found a monotonic
variation of the differential conductance and TMR  with the angle
between magnetic moments of the leads. This typical angular
dependence is due to the fact that different contributions to the
exchange field cancel each other and there is no spin precession
of the dot spin -- the dot spin remains in the plane defined by
the magnetic moments of the leads.

When the system is described by an asymmetric Anderson model, the
angular dependence of the differential conductance and TMR becomes
more complex. First, the maximum and minimum values of
differential conductance occur not for collinear configurations,
but when the leads' magnetizations are non-collinearly aligned.
Second, the conductance is enhanced in a broad range of the angle
between the magnetic moments of the leads, and drops to the
minimum value when the configuration becomes close to the
antiparallel one. The conductance drop is particularly pronounced
for small bias voltages. This behavior results from the interplay
between the first-order processes giving rise to the exchange
field and the second-order processes that drive the current. The
exchange field also leads to a nonmonotonic angular dependence of
TMR. For magnetic configurations where differential conductance is
enhanced as compared to that in parallel configuration, the TMR
becomes negative.

\begin{acknowledgments}

We acknowledge discussions with J\"urgen K\"onig and Matthias
Braun. The work was supported by the Ministry of Science and
Higher Education (Poland) as research project (2006-2009). I. W.
also acknowledges support from the Foundation for Polish Science.

\end{acknowledgments}

%%%%%%%%%%%%%%%%%%%%%%%%%%%%%%%%%%%%%%%%%%%%%%%%%%%%%%%%%%%%%%%%
%%%%%%%%%%%%%%%%%%%%%%%%%%%%%%%%%%%%%%%%%%%%%%%%%%%%%%%%%%%%%%%%

\appendix

\section{Self-energies in the Coulomb blockade regime}

In this appendix we present the analytical formulas for the first-
and second-order self-energies involving at the ends the
off-diagonal states that are necessary to determine the dot
occupations and spin from Eq.~(\ref{masterCB}). Our results are
valid in the case of deep Coulomb blockade,
$|\varepsilon|,|\varepsilon+U|\gg k_{\rm B}T,\Gamma$. For the
first order we find, ${\Sigma_{0\sigma}^{0\bar{\sigma}}}^{(1)} =
{\Sigma_{{\rm d}\sigma}^{{\rm d}\bar{\sigma}}}^{(1)} = 2\pi i
\sum_\alpha \Gamma_\alpha^{\sigma\bar{\sigma}}$,
${\Sigma_{\sigma\sigma}^{\sigma\bar{\sigma}}}^{(1)} =
{\Sigma_{\sigma\sigma}^{\bar{\sigma}\sigma}}^{(1)} = \sum_\alpha
\left( B_{1\alpha}^{\sigma\bar{\sigma}} -
C_{1\alpha}^{\sigma\bar{\sigma}}\right)$, and
${\Sigma^{\sigma\sigma}_{\bar{\sigma}\bar{\sigma}}}^{(1)} =
\sum_\alpha \left( B_{1\alpha}^{\sigma\sigma} -
B_{1\alpha}^{\bar{\sigma}\bar{\sigma}} +
C_{1\alpha}^{\bar{\sigma}\bar{\sigma}} -
C_{1\alpha}^{\sigma\sigma}\right)$, where $\alpha={\rm L,R}$.

The imaginary parts of the second-order self-energies involving
off-diagonal states are given by
\begin{eqnarray}
\lefteqn{
   {\rm Im} {\Sigma_{\sigma\sigma}^{\sigma\bar{\sigma}}}^{(2)}
     = \pi  \sum_{\alpha,\beta}\sum_{\sigma^\prime} \bigg[
      2 \Gamma_\alpha^{\sigma\bar{\sigma}} \left(
         B_{2\beta}^{\sigma \sigma} +
         C_{2\beta}^{\bar{\sigma} \bar{\sigma}}
       \right)  } \nonumber\\
       & + & 2 B_{2\alpha\beta}^{\sigma \sigma,\sigma \bar{\sigma}}
       - B_{2\alpha\beta}^{\sigma \bar{\sigma},\sigma^\prime \sigma^\prime}
       + 2 C_{2\alpha\beta}^{\sigma \bar{\sigma},\bar{\sigma} \bar{\sigma}}
       - C_{2\alpha\beta}^{\sigma^\prime \sigma^\prime,\sigma
       \bar{\sigma}}\nonumber\\
       & + &  \frac{1}{U}\left(
         A_{1\alpha\beta}^{\bar{\sigma} \bar{\sigma},\sigma \bar{\sigma}}
         - A_{1\alpha\beta}^{\sigma \bar{\sigma},\sigma \sigma}
         + 3A_{1\alpha\beta}^{\sigma \sigma, \sigma \bar{\sigma}}
         - 3A_{1\alpha\beta}^{\sigma \bar{\sigma},\bar{\sigma} \bar{\sigma}}
       \right)
     \bigg] \,,
\end{eqnarray}
\begin{eqnarray}
\lefteqn{
   {\rm Im} {\Sigma_{\sigma\sigma}^{\bar{\sigma}\sigma}}^{(2)}
     = \pi \sum_{\alpha,\beta} \bigg[
      2 \Gamma_\alpha^{\sigma\sigma} B_{2\beta}^{\sigma \bar{\sigma}} +
      2 \Gamma_\alpha^{\bar{\sigma}\bar{\sigma}} C_{2\beta}^{\sigma \bar{\sigma}}
        } \nonumber\\
       & + & B_{2\alpha\beta}^{\sigma \bar{\sigma},\sigma \sigma }
       - B_{2\alpha\beta}^{\sigma \bar{\sigma},\bar{\sigma}\bar{\sigma}}
       + C_{2\alpha\beta}^{\sigma \sigma,\sigma \bar{\sigma} }
       - C_{2\alpha\beta}^{\bar{\sigma}\bar{\sigma},\sigma\bar{\sigma}}
       \nonumber\\
       & + & \frac{1}{U}\left(
         A_{1\alpha\beta}^{\sigma \sigma, \sigma \bar{\sigma}}
       - A_{1\alpha\beta}^{\bar{\sigma} \bar{\sigma},\sigma \bar{\sigma}}
       + A_{1\alpha\beta}^{\sigma \bar{\sigma},\sigma \sigma}
       - A_{1\alpha\beta}^{\sigma \bar{\sigma},\bar{\sigma} \bar{\sigma}}
       \right)
     \bigg] \,,
\end{eqnarray}
\begin{eqnarray}
\lefteqn{
   {\rm Im} {\Sigma_{\bar{\sigma}\sigma}^{\sigma\bar{\sigma}}}^{(2)}
     = 2 \pi \sum_{\alpha,\beta} \bigg[
      \Gamma_\alpha^{\sigma\bar{\sigma}} B_{2\beta}^{\sigma \bar{\sigma}}
      - \Gamma_\alpha^{\sigma\bar{\sigma}} C_{2\beta}^{\sigma
      \bar{\sigma}}}
        \nonumber \\
       & + & B_{2\alpha\beta}^{\sigma \bar{\sigma},\sigma \bar{\sigma}}
       + C_{2\alpha\beta}^{\sigma \bar{\sigma},\sigma \bar{\sigma}}
       + \frac{2}{U} A_{1\alpha\beta}^{\sigma \bar{\sigma},\sigma \bar{\sigma}}
      \bigg] \,,
\end{eqnarray}
\begin{eqnarray}
\lefteqn{
   {\rm Im} {\Sigma_{\bar{\sigma}\bar{\sigma}}^{\sigma\sigma}}^{(2)}
     = \pi \sum_{\alpha,\beta}\sum_{\sigma^\prime,\sigma^{\prime\prime}}
     \bigg[
      2 \Gamma_\alpha^{\sigma\bar{\sigma}} \left(
         B_{2\beta}^{\sigma \bar{\sigma}} -
         C_{2\beta}^{\sigma \bar{\sigma}}
       \right)  } \nonumber\\
       & + & 2 B_{2\alpha\beta}^{\sigma \bar{\sigma},\sigma \bar{\sigma}}
       + 2 C_{2\alpha\beta}^{\sigma \bar{\sigma},\sigma \bar{\sigma}}
       - B_{2\alpha\beta}^{\sigma^\prime \sigma^\prime,
         \sigma^{\prime\prime}\sigma^{\prime\prime} }
       - C_{2\alpha\beta}^{\sigma^\prime \sigma^\prime,
         \sigma^{\prime\prime}\sigma^{\prime\prime} }
         \nonumber\\
       & + &  \frac{2}{U}\left(
        2 A_{1\alpha\beta}^{\sigma \bar{\sigma},\sigma \bar{\sigma}}
        - A_{1\alpha\beta}^{\sigma^\prime \sigma^\prime,
         \sigma^{\prime\prime}\sigma^{\prime\prime} }
       \right)
     \bigg] \,,
\end{eqnarray}
\begin{equation}
   {\rm Im} {\Sigma_{\bar{\sigma}0}^{\sigma 0}}^{(2)}
     = 2 \pi \sum_{\alpha,\beta}
        \left(\Gamma_\alpha^{\sigma\sigma} + \Gamma_\alpha^{\bar{\sigma}\bar{\sigma}} \right)
        B_{2\beta}^{\sigma \bar{\sigma}} \,,
\end{equation}
where $\alpha,\beta={\rm L,R}$ and $\sigma^\prime,
\sigma^{\prime\prime} =\sigma,\bar{\sigma}$. The corresponding
parameters are defined as
\begin{equation}\label{Eq:AppB1}
  B_{n\alpha}^{\sigma\sigma^\prime} = \Gamma_\alpha^{\sigma
  \sigma^\prime} X_n(\varepsilon-\mu_\alpha) \,,
\end{equation}
\begin{eqnarray}\label{Eq:AppB2}
  B_{n\alpha\beta}^{\sigma\sigma^\prime\sigma^{\prime\prime}\sigma^{\prime\prime\prime}}
  &=& \Gamma_\alpha^{\sigma \sigma^\prime}
  \Gamma_\beta^{\sigma^{\prime\prime}\sigma^{\prime\prime\prime}}
  f_{\rm B}(\mu_\alpha-\mu_\beta) \nonumber\\
  &\times& \left[
  X_n(\varepsilon-\mu_\alpha)-X_n(\varepsilon-\mu_\beta)\right] \,,
\end{eqnarray}
with
\begin{equation}
   X_{n+1}(x) =
   \frac{d^{(n)}}{dx^{(n)}}
   {\rm Re} \left[
     \Psi\left( \frac{1}{2} + i\frac{x}{2\pi k_{\rm
     B}T}\right) - \ln \left( \frac{W}{2\pi k_{\rm B}T} \right)
   \right] \,,
\end{equation}
where $W$ is the cutoff parameter and $f_{\rm B}$ denotes the Bose
function. The parameters $C_{n\alpha}^{\sigma\sigma^\prime}$ and
$C_{n\alpha\beta}^{\sigma\sigma^\prime
\sigma^{\prime\prime}\sigma^{\prime\prime\prime}}$ are given by
Eqs.~(\ref{Eq:AppB1}) and (\ref{Eq:AppB2}), respectively, with
$\varepsilon$ replaced by $\varepsilon+U$, while $
A_{n\alpha\beta} ^{\sigma\sigma^\prime\sigma ^{\prime\prime}
\sigma^{\prime\prime\prime}} = B_{n\alpha\beta}^
{\sigma\sigma^\prime\sigma ^{\prime\prime}
\sigma^{\prime\prime\prime}} - C_{n\alpha\beta}
^{\sigma\sigma^\prime\sigma ^{\prime\prime}\sigma
^{\prime\prime\prime}}$. The other self-energies can be found
using the rules and relations given in section \ref{sec:3a}.

\section{Green's functions in the Coulomb blockade regime}

In the following we present the explicit formulas for the Green's
functions needed to calculate the current in the case of the
Coulomb blockade regime. The Green's functions
$G_{\sigma\sigma^\prime}^{\lessgtr (1,0)}$ are given by
$G_{\sigma\sigma}^{> (1,0)} = - 2\pi i
{P_0^0}^{(1)}\delta(\omega-\varepsilon)$, $G_{\sigma\sigma}^{<
(1,0)} = 2\pi i {P_{\rm d}^{\rm d}}^{(1)}
\delta(\omega-\varepsilon-U)$, while
$G_{\sigma\bar{\sigma}}^{\lessgtr (1,0)} = 0$. On the other hand,
for $G_{\sigma\sigma^\prime}^{\lessgtr (0,1)}$ we find
\begin{eqnarray}
   G_{\sigma\sigma}^{> (0,1)} &=& - 2\pi i \sum_{\sigma^\prime} \left\{
      {P_{\sigma^\prime}^{\sigma^\prime}}^{(0)} \left[
         b_{2\alpha}^{-\sigma^\prime \sigma^\prime}
         + B_{2\alpha}^{\sigma^\prime \sigma^\prime}  \delta(\omega-\varepsilon)
       \right] \right. \nonumber\\
       && + {P_{\bar{\sigma}}^{\bar{\sigma}}}^{(0)} \left[
         c_{2\alpha}^{-\sigma^\prime \sigma^\prime}
         + \frac{2}{U} \left(
           b_{1\alpha}^{-\bar{\sigma} \bar{\sigma}} -
           c_{1\alpha}^{-\bar{\sigma} \bar{\sigma}}
         \right)
       \right] \nonumber\\
       && + 2 S_x^{(0)} \left[
         b_{2\alpha}^{-\sigma\bar{\sigma}}
         + B_{2\alpha}^{\sigma\bar{\sigma}}\delta(\omega-\varepsilon)
         \right. \nonumber\\
         && \left.\left.+ \frac{1}{U}\left(
           b_{1\alpha}^{-\sigma\bar{\sigma}} -
           c_{1\alpha}^{-\sigma\bar{\sigma}}
         \right)
       \right]
     \right\} \,,
\end{eqnarray}
\begin{eqnarray}
   G_{\sigma\bar{\sigma}}^{> (0,1)} &=& 2\pi i \left\{
      S_x^{(0)} \sum_{\sigma^\prime} \left[
         c_{2\alpha}^{-\sigma^\prime \sigma^\prime}
         + \frac{1}{U} \left(
           b_{1\alpha}^{-\sigma^\prime \sigma^\prime} -
           c_{1\alpha}^{-\sigma^\prime \sigma^\prime}
         \right)
       \right]\right. \nonumber\\
       && \left.+ \frac{1}{U}\left[
           {P_{\sigma}^{\sigma}}^{(0)}
           + {P_{\bar{\sigma}}^{\bar{\sigma}}}^{(0)}
          \right]
          \left(
            b_{1\alpha}^{-\sigma\bar{\sigma}} -
            c_{1\alpha}^{-\sigma\bar{\sigma}}
          \right)
     \right\} \,,
\end{eqnarray}
\begin{eqnarray}
   G_{\sigma\sigma}^{< (0,1)} &=& 2\pi i \sum_{\sigma^\prime} \left\{
      {P_{\bar{\sigma}^\prime}^{\bar{\sigma}^\prime}}^{(0)} \left[
         c_{2\alpha}^{+\sigma^\prime \sigma^\prime}
         - C_{2\alpha}^{\sigma^\prime \sigma^\prime}  \delta(\omega-\varepsilon-U)
       \right] \right. \nonumber\\
       && + {P_{\sigma}^{\sigma}}^{(0)} \left[
         b_{2\alpha}^{+\sigma^\prime \sigma^\prime}
         + \frac{2}{U} \left(
           b_{1\alpha}^{+\bar{\sigma} \bar{\sigma}} -
           c_{1\alpha}^{+\bar{\sigma} \bar{\sigma}}
         \right)
       \right] \nonumber\\
       && - 2 S_x^{(0)} \left[
         c_{2\alpha}^{+\sigma\bar{\sigma}}
         - C_{2\alpha}^{\sigma\bar{\sigma}}\delta(\omega-\varepsilon-U)
         \right. \nonumber\\
         && \left.\left.+ \frac{1}{U}\left(
           b_{1\alpha}^{+\sigma\bar{\sigma}} -
           c_{1\alpha}^{+\sigma\bar{\sigma}}
         \right)
       \right]
     \right\} \,,
\end{eqnarray}
\begin{eqnarray}
   G_{\sigma\bar{\sigma}}^{< (0,1)} &=& 2\pi i \left\{
      S_x^{(0)} \sum_{\sigma^\prime} \left[
         b_{2\alpha}^{+\sigma^\prime \sigma^\prime}
         + \frac{1}{U} \left(
           b_{1\alpha}^{+\sigma^\prime \sigma^\prime} -
           c_{1\alpha}^{+\sigma^\prime \sigma^\prime}
         \right)
       \right]\right. \nonumber\\
       && \left.- \frac{1}{U}\left[
           {P_{\sigma}^{\sigma}}^{(0)}
           + {P_{\bar{\sigma}}^{\bar{\sigma}}}^{(0)}
          \right]
          \left(
            b_{1\alpha}^{+\sigma\bar{\sigma}} -
            c_{1\alpha}^{+\sigma\bar{\sigma}}
          \right)
     \right\} \,,
\end{eqnarray}
with
\begin{equation}\label{Eq:AppB4}
   b_{n\alpha}^{\pm \sigma\sigma^\prime} =
   \frac{\gamma_\alpha^{\pm\sigma
   \sigma^\prime}(\omega)}{(\omega-\varepsilon)^n}\,,
\end{equation}
and $c_{n\alpha}^{\pm \sigma\sigma^\prime}$ is given by
Eq.~(\ref{Eq:AppB4}) with $\varepsilon\leftrightarrow
\varepsilon+U$, while the other parameters are defined in Appendix
A. In the above formulas we have cancelled all the terms whose
contribution to the current is exponentially suppressed, i.e., the
terms multiplied with $\delta(\omega-\varepsilon)$ and
$\delta(\omega-\varepsilon-U)$ in $G_{\sigma\sigma^\prime}^{<
(0,1)}$ and $G_{\sigma\sigma^\prime}^{> (0,1)}$, respectively, as
well as the terms multiplied with
$\delta^\prime(\omega-\varepsilon)$ and
$\delta^\prime(\omega-\varepsilon-U)$, where $\delta$ is the Dirac
function and $\delta^\prime$ its derivative.

%%%%%%%%%%%%%%%%%%%%%%%%%%%%%%%%%%%%%%%%%%%%%%%%%%%%%%%%%%%%%%%%
%%%%%%%%%%%%%%%%%%%%%%%%%%%%%%%%%%%%%%%%%%%%%%%%%%%%%%%%%%%%%%%%

\end{document}